\documentclass[10pt, onecolumn, conference]{IEEEtran}

\usepackage{enumitem}
\usepackage{amsmath} 

\usepackage{booktabs,array}
\usepackage{graphicx,url,tabularx}
\usepackage[brazil]{babel}
\usepackage[utf8]{inputenc}  
\usepackage{ragged2e}
\usepackage{comment}
\usepackage{ifthen}
\usepackage{graphicx}
\usepackage{tcolorbox}
\usepackage{amssymb}
\usepackage{pifont}
\usepackage[table]{xcolor}

\usepackage{enumerate}

\newif\iffinal
\finalfalse

\title{\emph{NetSecBed}: A Container-Native Testbed for Reproducible Cybersecurity Experimentation}

\author{
\IEEEauthorblockN{
    Leonardo Bitzki\IEEEauthorrefmark{1}\IEEEauthorrefmark{2},
    Diego Kreutz\IEEEauthorrefmark{2},
    Tiago Heinrich\IEEEauthorrefmark{3},\\
    Douglas Fideles\IEEEauthorrefmark{2},
    Leandro Bertholdo\IEEEauthorrefmark{1},
    Silvio Quincozes\IEEEauthorrefmark{2},
    Angelo Diniz\IEEEauthorrefmark{2}
}
\IEEEauthorblockA{\IEEEauthorrefmark{1}Federal University of Rio Grande do Sul (UFRGS)}
\IEEEauthorblockA{\IEEEauthorrefmark{2}AI Horizon Labs and PPGES -- Federal University of Pampa (UNIPAMPA)}
\IEEEauthorblockA{\IEEEauthorrefmark{3}Max Planck Institute for Informatics (MPI-INF)}
}

\begin{document} 
\maketitle

\begin{abstract}
Cybersecurity research increasingly depends on reproducible evidence, such as traffic traces, logs, and labeled datasets, yet most public datasets remain static and offer limited support for controlled re-execution and traceability, especially in heterogeneous multi-protocol environments. This paper presents \textit{NetSecBed}, a container-native, scenario-oriented testbed for reproducible generation of network traffic evidence and execution artifacts under controlled conditions, particularly suitable for IoT, IIoT, and pervasive multi-protocol environments. The framework integrates 60 attack scenarios, 9 target services, and benign traffic generators as single-purpose containers, enabling plug-and-play extensibility and traceability through declarative specifications. Its pipeline automates parametrized execution, packet capture, log collection, service probing, feature extraction, and dataset consolidation. The main contribution is a repeatable, auditable, and extensible framework for cybersecurity experimentation that reduces operational bias and supports continuous dataset generation.
\end{abstract}
     
\section{Introduction}
\label{sec:introduction}

Experimental cybersecurity research increasingly depends on reproducible evidence, such as traffic traces, logs, telemetry, and labeled datasets, to enable rigorous and comparable evaluation of detection, classification, and incident response mechanisms. Yet generating such evidence remains an open technical and methodological challenge. In modern environments, experimental fidelity is tightly coupled with ecosystem heterogeneity, spanning traditional services (e.g., HTTP, SSH, and SMB), IoT protocols (e.g., MQTT and CoAP), and emerging communication frameworks (e.g., Zenoh and XRCE-DDS), where even minor variations in software versions, timing, configuration, and background noise may significantly alter observed artifacts and compromise cross-experiment comparability. This creates a recurring trade-off between realism, operational cost, and reproducibility: higher realism typically increases deployment, instrumentation, and auditing complexity.

This challenge is further aggravated by the growing obsolescence of public cybersecurity datasets. Continuous changes in service behavior, protocol stacks, and adversarial strategies progressively reduce the temporal validity of static collections, limiting both scientific evaluation and model training. Consequently, the central problem is no longer merely how to obtain data, but how to systematically generate experimental evidence in a repeatable, auditable, and extensible manner, preserving full traceability of origin, parameters, temporal context, and execution conditions.

The state of the art remains fragmented. Existing testbeds often optimize isolated dimensions, such as physical realism, protocol specificity, low-cost deployment, or fidelity in specialized domains (e.g., IoT, IIoT, SCADA, and CPS), while still exhibiting recurring limitations in \textit{(i)} attack coverage, \textit{(ii)} protocol heterogeneity, \textit{(iii)} experimental scalability, and \textit{(iv)} support for systematic re-execution and continuous dataset generation. In practice, many approaches still rely on static collections, domain-coupled scenarios, or partially manual pipelines, which hinder comparability, auditing, and incremental evolution of the experimental environment.

To address these limitations, we propose \textit{NetSecBed}\footnote{\textit{NetSecBed} is available at \url{https://github.com/ANONIMIZADO_PARA_REVISAO}.}, a container-native, scenario-oriented, and natively extensible testbed for reproducible evidence generation in heterogeneous network environments. The current architecture integrates 60 attack scenarios, 9 target services, and parameterizable benign traffic models, all encapsulated as single-purpose containers and described through declarative metadata specifications. Unlike approaches centered on static datasets, \textit{NetSecBed} provides a complete and auditable experimental pipeline, including scenario selection, parametrized execution, packet capture, log collection, service probing, feature extraction, and automatic dataset consolidation. 

Our central methodological hypothesis is that explicitly separating declarative scenario specification from orchestrated execution reduces operational variability, improves traceability, and enables incremental scalability of the attack catalog without coupling extensions to the system core. This design allows new attacks, services, and benign profiles to be incorporated through YAML specifications and containerized artifacts, supporting continuous sample renewal and improving the external validity of experimental results. This makes \textit{NetSecBed} particularly relevant for cybersecurity experimentation in ubiquitous and pervasive computing environments, where heterogeneous devices, protocols, and continuously evolving traffic patterns are intrinsic characteristics.

The main contributions of this work are: \textit{(i)} an observable, container-native architecture for reproducible cybersecurity experimentation; \textit{(ii)} a structured and extensible catalog of 60 attacks with support for multiple protocols and services; \textit{(iii)} an automated pipeline for continuous artifact and dataset generation; and \textit{(iv)} a methodological framework that reduces operational bias and improves comparability across heterogeneous experiments.

\section{Related Work} \label{sec:trabalhos_relacionados}

Recent cybersecurity testbeds span a broad spectrum of objectives, ranging from IoT vulnerability assessment and traffic generation to cyber-physical attack impact analysis. However, as summarized in Table~\ref{tab:testbed_comparison}, the current state of the art remains fragmented across four critical methodological dimensions: scalability, protocol heterogeneity, attack coverage, and reproducibility.

\begin{table*}[!htp]
\centering
\small
\caption{State-of-the-art comparison of cybersecurity testbeds.}
\label{tab:testbed_comparison}
\setlength{\tabcolsep}{3pt}
\renewcommand{\arraystretch}{1.15}
\resizebox{\textwidth}{!}{
\begin{tabular}{p{2.4cm} p{4.0cm} p{2.4cm} p{2.4cm} c p{2.8cm} p{4.5cm}}
\toprule
\textbf{Testbed} & \textbf{Scientific Contributions} & \textbf{Scalability} & \textbf{Protocols} & \textbf{\#Atk} & \textbf{Target Services} & \textbf{Main Limitations} \\
\midrule

\textbf{Gotham} &
Reproducible IoT testbed; multi-protocol support; real dataset generation &
$\sim$140 nodes / machine &
MQTT, CoAP, RTSP &
18--25 &
MQTT broker, CoAP, RTSP &
Restricted service scope; limited device diversity; no  MITRE ATT\&CK mapping \\

\textbf{RESLab} &
Security experimentation and dataset collection &
$\sim$50 substations &
DNP3 &
4 &
DNP3 masters / outstations &
Very limited attack diversity; single-protocol scope; no MITRE mapping \\

\textbf{IoT Security Testbed} &
Open-source automated vulnerability assessment platform for IoT devices &
Low (1 device / run) &
HTTP(S), UDP, FTP, SSH, Telnet &
13 &
Camera, smart bulb, IoT web services &
Few attacks; minimal scalability; no native IoT protocols or MITRE mapping \\

\textbf{LoRaWAN MitM} &
LPWAN security testbed for real over-the-air experiments &
Very low &
LoRaWAN 1.0.2 &
2 &
LoRa end devices, gateway &
Very limited attack coverage; radio-dependent; unrealistic assumptions \\

\textbf{CPS Microgrid} &
Real-time co-simulation for CPS attack impact analysis &
Very low &
Modbus, IEC 61850, DNP3 &
3 &
IEDs, RTUs, PMUs &
Low scalability; protocol limitations; proprietary hardware dependence \\

\textbf{QR-GridEx} &
HIL-based large-scale coordinated attack simulation for smart grids &
Medium &
DNP3, IEC 61850, C37.118 &
10 &
SCADA/WAMS, substations &
Limited attack catalog; low reproducibility; energy-specific scope \\

\textbf{Lightweight SCADA} &
Traffic generation and attack collection for SCADA security datasets &
Lightweight &
Modbus &
2 &
SCADA-LTS, Conpot &
Only two attacks; single-protocol scope; no quantitative metrics or MITRE mapping \\

\textbf{NetSecBed (ours)} &
Reproducible attack pipeline; structured attack catalog; feature extraction; observability and extensibility by design &
High (container-native) &
Web, MySQL, SSH, Telnet, SMB, MQTT, CoAP, Zenoh, XRCE-DDS &
60 &
Web, DB, SSH, SMB, IoT/IIoT brokers &
External threat focus; no firmware/CVE-specific or industrial protocol attacks \\

\bottomrule
\end{tabular}
}
\end{table*}

Recent survey studies have reinforced the role of cybersecurity testbeds as scientific infrastructures for reproducible experimentation, dataset generation, and systematic benchmarking. In IoT and IIoT domains, recent reviews consistently highlight persistent challenges in scalability, protocol heterogeneity, reproducibility, and experimental realism~\cite{de2024cybersecurity, akinremi2025systematic}. Similar limitations have also been reported in evaluation frameworks for anomaly detection in built environments, further emphasizing the need for large-scale, repeatable, and well-instrumented infrastructures for trustworthy security research~\cite{alosaimi2025testbeds}.

Within this landscape, representative testbeds typically optimize specific methodological dimensions. In the IoT domain, the \textit{Gotham Testbed}~\cite{saez2023gotham} provides a reproducible architecture with multi-protocol support (MQTT, CoAP, and RTSP) and moderate scalability, reaching approximately 140 nodes per machine. However, it remains restricted to a narrow set of IoT services, limited device diversity, and lacks explicit alignment with standardized adversarial taxonomies such as MITRE ATT\&CK. Similarly, the \textit{IoT Security Testbed}~\cite{siboni2018security} advances automated vulnerability assessment for real devices, but offers only 13 attacks, low scalability, and no support for native IoT messaging protocols such as MQTT or CoAP, limiting its representativeness in modern IoT/IIoT scenarios.

A second line of work focuses on industrial control systems and SCADA experimentation. \textit{RESLab}~\cite{wlazlo2021man} and the \textit{Lightweight SCADA Testbed}~\cite{khan2020lightweight} provide relevant infrastructures for traffic generation and dataset construction, yet their methodological breadth is constrained by single-protocol designs, very limited attack catalogs, and reduced cross-domain applicability. Likewise, cyber-physical platforms such as the \textit{CPS Microgrid Testbed}~\cite{cPSMicrogrid}, \textit{QR-GridEx}~\cite{qRGridEx}, and the \textit{LoRaWAN MitM Testbed}~\cite{loRaWAN} improve realism through co-simulation, HIL, and hardware coupling, but remain strongly domain-specific, infrastructure-dependent, and difficult to generalize.

Taken together, these works reveal a recurring trade-off: prior testbeds typically improve realism, protocol specificity, or lightweight deployment at the expense of attack breadth, protocol diversity, and reproducibility. In contrast, \textbf{NetSecBed} is explicitly designed as a scientifically reproducible and extensible benchmarking framework for large-scale cybersecurity experimentation. Its main contributions include a container-native architecture, a structured catalog of 60 attacks, support for heterogeneous traditional and IoT/IIoT protocols (e.g., MQTT, CoAP, Zenoh, and XRCE-DDS), integrated \textit{pcap}/log collection, automated feature extraction, and observability by design. These characteristics enable repeatable experimentation, continuous dataset generation, controlled comparative evaluation, and higher external validity, substantially advancing the state of the art in scalability, reproducibility, and experimental breadth.

\section{NetSecBed Architecture}
\label{sec:architecture}

The core scientific contribution of \textit{NetSecBed} lies in its architecture-first methodology for reproducible cybersecurity experimentation. Rather than treating attack execution as an isolated operational task, the proposed architecture formalizes experimentation as a controlled, observable, and repeatable pipeline capable of generating traceable evidence artifacts, including traffic captures, logs, telemetry, metrics, datasets, and execution reports. This architectural design directly addresses the central limitations identified in the state of the art, namely limited reproducibility, weak traceability, fragmented pipelines, and poor extensibility of attack catalogs.

To achieve this, the architecture is explicitly designed around four methodological principles: \textit{(i)} strict component isolation, \textit{(ii)} standardized experimental execution, \textit{(iii)} declarative extensibility of attack scenarios, and \textit{(iv)} end-to-end traceability of generated artifacts. Together, these principles enable systematic re-execution of experiments under controlled conditions while reducing operational bias and ensuring comparability across multiple runs and scenarios.

Within the scope of this work, \textit{NetSecBed} assumes a container-native experimental environment in which attackers, target services, and benign clients are instantiated and fully controlled by the testbed itself. The focus is on network-centric attack scenarios executed under controlled conditions, with explicit observation of traffic dynamics, service-level telemetry, and operational effects over instrumented services. The objective is not to fully replicate the complexity of production enterprise environments, but rather to provide a scientifically controlled infrastructure for repeatable experimentation. Accordingly, threats outside the scope of experimental execution, such as physical compromise of the host infrastructure or persistence beyond container boundaries, are intentionally excluded from the threat model.

From an architectural perspective, \textit{NetSecBed} is organized into four complementary subsystems: \textit{(i)} a containerized artifact repository, comprising attackers, target services, and benign traffic generators; \textit{(ii)} an experiment operation and configuration layer; \textit{(iii)} an orchestration and instrumentation layer, responsible for transforming attack execution into a complete experimental cycle that produces verifiable and comparable artifacts; and \textit{(iv)} a data acquisition and processing layer, responsible for capture, collection, feature extraction, transformation, and generation of datasets, visual reports, and execution evidence.

\begin{figure}[!ht]
    \centering
    \includegraphics[width=0.9\textwidth]{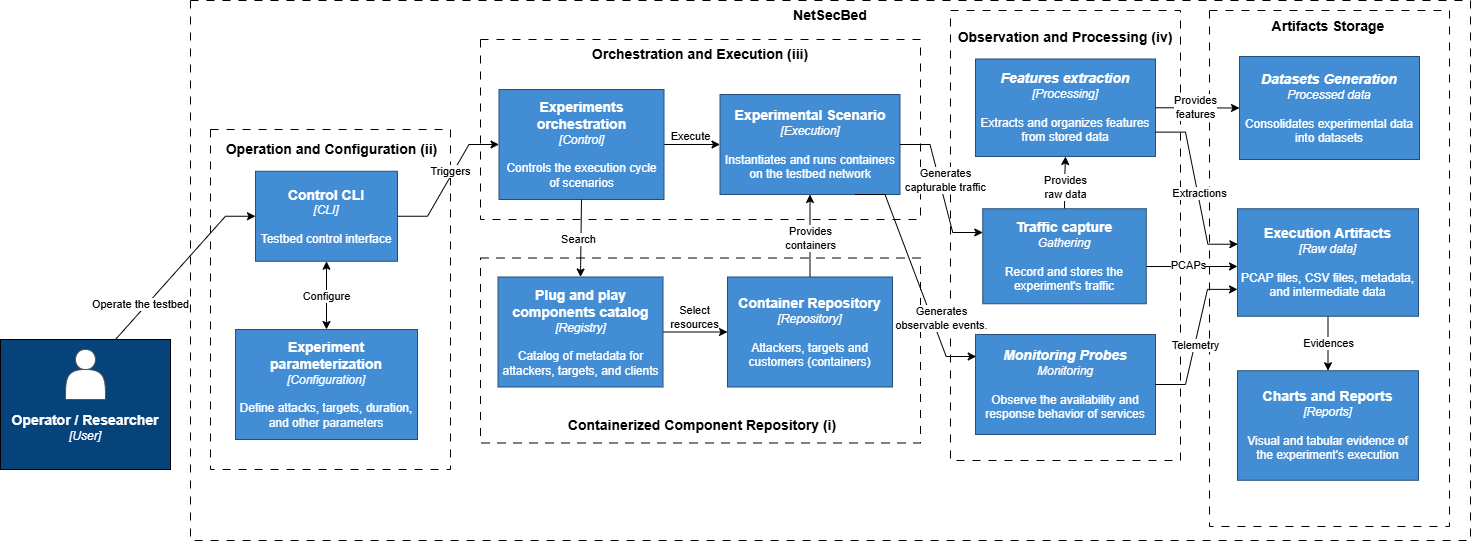}
    \caption{Overview of the \textit{NetSecBed} architecture.}
    \label{fig:architecture}
\end{figure}

Figure~\ref{fig:architecture} provides a consolidated view of the proposed architecture. At a high level, the operator defines the experimental parameters through the control interface, while the orchestrator instantiates the required scenario components from the container repository and executes the experiment according to a controlled lifecycle. This lifecycle produces observable traffic and service events in the experimental network, which are continuously captured and transformed into structured features. These outputs are subsequently consolidated into higher-level artifacts, including datasets, visual analytics, execution reports, and reproducible evidence traces.

The key architectural innovation is the explicit decoupling between scenario specification, execution orchestration, and artifact generation. This separation transforms the testbed from a simple attack execution platform into a reproducible scientific framework for evidence generation, enabling scalable expansion of attack catalogs and systematic regeneration of up-to-date datasets as services, tools, and network conditions evolve.

\subsection{Containerized Artifact Repository}
\label{subsec:artifact_repository}

A central methodological contribution of \textit{NetSecBed} is its containerized artifact repository, which formalizes each attack scenario as an isolated, single-purpose experimental component. In this design, every attack is encapsulated as an independent container, enforcing strict execution isolation, reducing ambiguity, and improving experimental traceability. Rather than relying on loosely coupled scripts or operator-dependent procedures, each scenario is modeled as a controlled composition of an attacker, one or more target services, optional benign clients, and the execution parameters that define the experimental conditions. This architectural decision is fundamental to reproducibility: by enforcing one objective per container, each experiment becomes fully auditable, allowing precise identification of what was executed, for how long, against which target, and under which parameters, while minimizing coupling and operational variability across repeated runs.

Beyond attacker containers, \textit{NetSecBed} includes a heterogeneous repository of target service containers, including web servers, MySQL, SSH, Telnet, SMB, MQTT, CoAP, Zenoh, and XRCE-DDS services, as well as parameterizable benign traffic generators capable of producing deterministic background traffic (e.g., service type, access frequency, inter-arrival interval, and total duration). This heterogeneity is a key scientific differentiator, enabling systematic experimentation across both traditional and IoT/IIoT communication ecosystems.

Each scenario is formally described through structured metadata, including identifier, description, image reference, supported parameters, operational and security notes, and explicit mappings to MITRE ATT\&CK tactics and techniques. Importantly, these metadata act as an \textit{experimental contract}, i.e., the minimum declarative specification required to reproduce the scenario and correctly interpret the generated artifacts, which is one of the main mechanisms through which \textit{NetSecBed} ensures traceability and long-term reproducibility. Furthermore, artifacts are semantically organized according to MITRE ATT\&CK~\cite{roy2023sok}, linking each implementation to both the adversarial objective (\textit{why}) and its operational mechanism (\textit{how}), which improves consistency, extensibility, and comparability across scenarios.

\noindent
\textbf{Adding new attacks.}
Another important contribution is the declarative extensibility model. New attacks are incorporated through three lightweight steps: \textit{(i)} creating a dedicated directory within \texttt{docker/}, where a \texttt{Dockerfile} defines the base image, dependencies, scripts, and execution environment; \textit{(ii)} encapsulating the attack logic within an \texttt{entrypoint.sh} hook; and \textit{(iii)} registering the artifact in \texttt{attack.yaml}, which connects it to the orchestration layer. This design allows the attack catalog to scale incrementally without modifying the system core, directly addressing a key limitation of prior testbeds: poor extensibility and high maintenance overhead.

\subsection{Experiment Operation and Orchestration Engine}
\label{subsec:orchestration_engine}

A key scientific contribution of \textit{NetSecBed} is its orchestration engine, which transforms an abstract experiment specification into an executable, observable, and fully repeatable workflow. Rather than executing attacks as isolated scripts, this layer formalizes experimentation as a controlled lifecycle, reducing operator-dependent variability and enabling systematic comparison across repeated runs. In practice, the engine receives parameters such as target service, attack set, intensity levels, repetitions, capture interface, traffic filters, and phase duration, and converts them into a structured execution matrix organized by service, attack, level, and repetition, supporting controlled sensitivity and comparative analyses.

\begin{figure}[!ht]
    \centering
    \includegraphics[width=\textwidth]{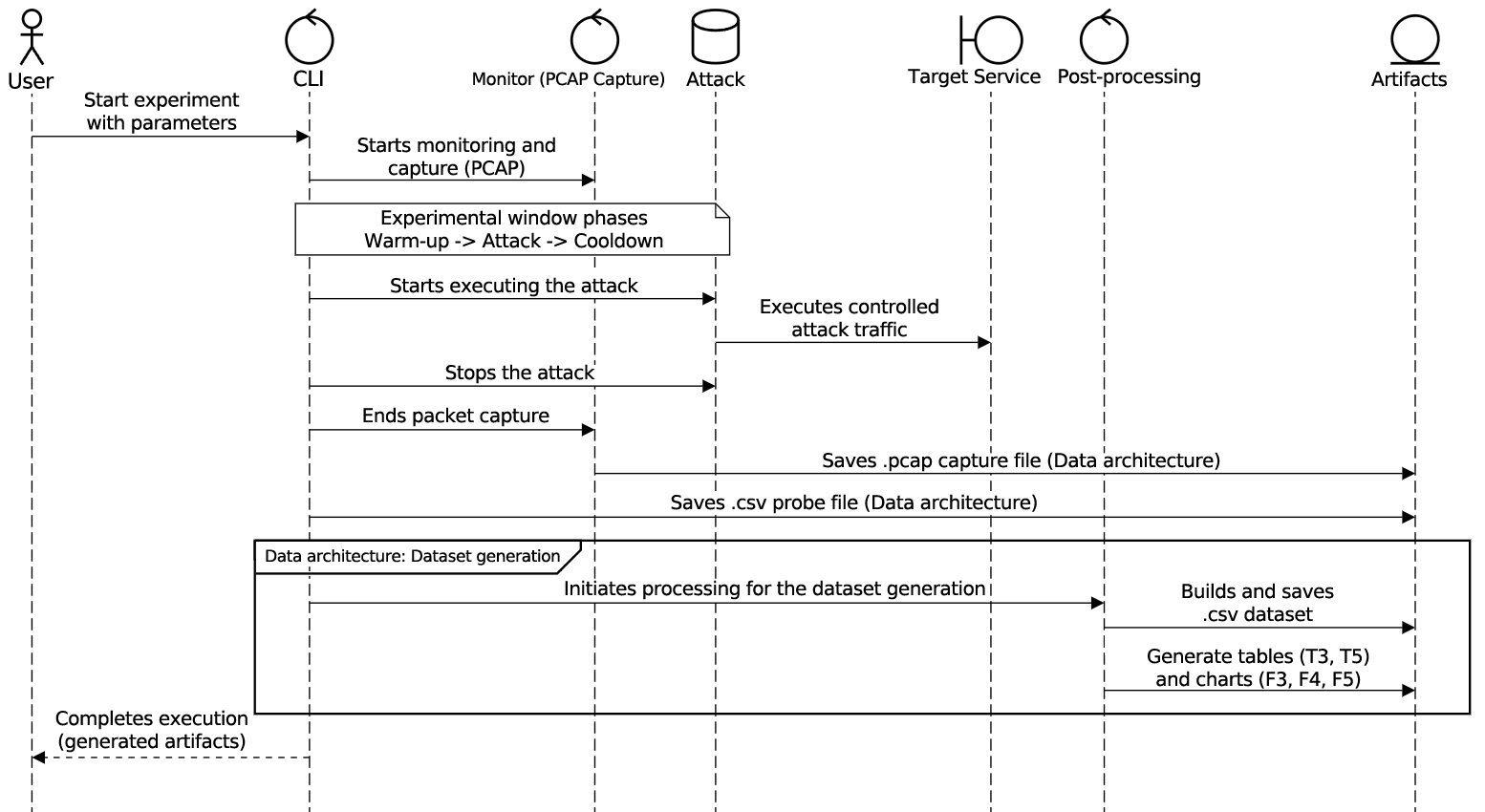}
    \caption{Execution pipeline of \textit{NetSecBed} for artifact generation.}
    \label{fig:diagrama_execucao}
\end{figure}

Operationally, orchestration follows an explicit \textit{warmup--attack--cooldown} lifecycle, separating service stabilization, adversarial interference, and recovery phases. In parallel, the engine continuously performs availability and latency probes over instrumented services, labeling each sample according to its temporal phase. This transforms each execution into a contextualized time series suitable for impact analysis, temporal robustness studies, and reproducible metric generation, as illustrated in Figure~\ref{fig:diagrama_execucao}. To preserve extensibility, attack execution is delegated to parameterized external \textit{hooks}, decoupling offensive logic from the system core and allowing new triggering mechanisms without redesigning the pipeline. Each run also produces a dedicated directory containing raw traffic captures, probe records, and metadata describing contextual parameters and execution timestamps, ensuring traceability and exact replay.

\subsection{Data Processing and Artifact Generation}
\label{subsec:data_processing_artifacts}

The data processing layer converts raw execution signals into analytically usable evidence. Beyond packet captures, it incorporates service probing time series and execution metadata, preserving the full context of each run. Feature extraction is performed through multiple independent mechanisms, including TShark~\cite{Wireshark}, Scapy~\cite{scapy.net}, and NTLFlowLyzer~\cite{SHAFI2025104160}, while maintaining separate extraction tracks prior to consolidation. This is a deliberate methodological choice: different extractors operate at distinct abstraction levels and protocol coverage, introducing different biases and sensitivities.

By preserving parallel extraction pipelines, \textit{NetSecBed} enables consistency verification, information-loss analysis, and robustness assessment of feature representations. The extracted outputs are then consolidated into standardized tabular datasets explicitly linked to execution metadata, enabling re-analysis, auditing, reproducibility, downstream detection tasks, and parameter sensitivity studies. In addition, the framework generates synthetic reports and visual analytics, including summary tables, latency metrics, failure-aware observations, temporal evolution, and cumulative distributions. This layer therefore transforms raw observations into traceable and continuously renewable evidence artifacts, directly supporting the generation of up-to-date cybersecurity datasets.

\subsection{Development and Experimental Environment}
\label{subsec:environment}

\textit{NetSecBed} is designed for lightweight deployment in both bare-metal and virtualized environments, prioritizing low replication cost, automated installation, and open-source components to improve reproducibility and facilitate adoption in research and teaching settings. The framework was developed and validated on commodity hardware in two representative environments: \textit{(i)} Ubuntu 24.04 under WSL on a Windows 10 host with an Intel Core i7-7700HQ, 16\,GB RAM, GTX 1060, and SSD storage, and \textit{(ii)} Kubuntu 24 LTS in bare-metal mode on an AMD Ryzen 5 5600X, 32\,GB RAM, RTX 3070 Ti, and NVMe storage, demonstrating that the proposed architecture can be reliably reproduced without specialized infrastructure.

\section{Methodology}
\label{sec:methodology}

Methodologically, \textit{NetSecBed} is formalized as a controlled evidence generation framework rather than a simple attack execution platform. Under this perspective, the focus extends beyond attack triggering to the experimental design itself, explicitly defining \textit{(i)} environmental control variables, \textit{(ii)} independent variables such as attack and benign traffic parameters, \textit{(iii)} dependent variables expressed as observable metrics, \textit{(iv)} the repetition plan, and \textit{(v)} acceptance criteria used to verify experimental consistency and artifact validity.

Each experiment is modeled as a tuple $\mathbf{E=(C,P,B,I,R)}$, where $\mathbf{C}$ represents the scenario (e.g., attack, target, and minimal topology), $\mathbf{P}$ defines the parameter space (e.g., rate, duration, port, and target), $\mathbf{B}$ describes the benign baseline (e.g., traffic profile, number of clients, and intervals), $\mathbf{I}$ specifies the instrumentation plan (e.g., \textit{pcap}, logs, and artifact generation), and $\mathbf{R}$ determines the repetition plan (e.g., number of runs and warmup/cooldown windows). This formalization improves traceability between configuration and results while enabling consistent comparison across scenarios and studies.

Operationally, experiments are executed in multiple rounds and explicitly segmented into \textit{warmup}, \textit{attack}, and \textit{cooldown} phases. The instrumentation plan includes availability probes, latency, error rate, relative time, packet capture, optional feature extraction, and automatic generation of analytical artifacts such as time series, latency CDFs, and failure-rate curves. This design enables methodological scalability, as distinct scenarios can be compared through homogeneous and reproducible metrics.

Among the evaluated metrics, latency percentiles p50, p95, and p99 are explicitly considered. While p50 captures the typical system response, p95 and p99 characterize tail behavior, exposing degradation in the worst 5\% and 1\% of requests, respectively. This choice is particularly suitable for cybersecurity experiments, as availability attacks tend to disproportionately affect distribution tails before significantly shifting the median~\cite{yang2025simulation}.

A key methodological principle is the explicit separation between execution and interpretation. The pipeline preserves both raw artifacts (\textit{pcaps} and logs) and derived artifacts (extracted features, consolidated datasets, tables, and figures), maintaining a complete evidence trail that supports reproducibility and auditing. Scientific analyses, including detection, classification, and metric computation, remain fully re-executable from the raw evidence, reducing dependence on \textit{ad hoc} transformations and minimizing the risk of conclusions based on irreproducible intermediate steps.

Finally, threats to validity are explicitly considered, including uncontrolled variables related to hardware noise, CPU/IO contention, host-level network interference, and sufficiency of the repetition plan, all of which may affect measurement stability and inference robustness.

\section{Evaluation}
\label{sec:evaluation}

To demonstrate the evidence-generation capabilities of \textit{NetSecBed}, we selected the \texttt{DoS SYN Flood} attack from the catalog of 60 implemented scenarios. This attack was chosen because it produces an immediate and measurable impact on target service availability, making it well suited to illustrate the relationship between attack intensity, duration, and disturbance onset. This design aligns with prior evaluation strategies~\cite{BERNIERI201786, 8565917, dharini2026efficient}, in which attack parameters and execution windows are systematically varied to analyze changes in system behavior, including network utilization, attack duration, and timing effects.

For the experimental analysis, we varied the \textit{rate} parameter controlling attack intensity together with the \textit{warmup}, \textit{attack}, and \textit{cooldown} windows that define the temporal structure of the experiment. We focus on levels L0 and L3, representing the lower and upper extremes of the adopted intensity scale, corresponding to up to 100 pps at L0 and an unrestricted rate at L3. This choice was designed to explicitly capture the contrast between a mild perturbation and a severe availability degradation scenario, while intermediate levels follow the same parametric logic and are omitted for space reasons.

Table~\ref{tab:ataques_t3_censored} summarizes the results. Under L0, the HTTP service remains stable across all phases, sustaining 100\% success rate and low latency percentiles, which confirms the baseline robustness of the target service under low-intensity conditions. In contrast, L3 produces a substantial degradation during the attack phase, reducing service success to 55.6\% and increasing latency from single-digit milliseconds to p50 = 1275.74\,ms and p95/p99 = 2002.2\,ms. Importantly, both warmup and cooldown phases preserve near-baseline behavior, which evidences that the observed degradation is temporally localized and directly attributable to the injected attack. This result demonstrates the ability of \textit{NetSecBed} to produce controlled, phase-aware, and quantitatively reproducible evidence of service degradation under adversarial conditions.

\begin{table}[!ht]
    \centering
    \footnotesize
    \resizebox{0.8\columnwidth}{!}{%
    \renewcommand{\arraystretch}{1.3}
    \begin{tabular}{cccccccc}
    \toprule
    \textbf{Level} & \textbf{Phase} & \textbf{Samples} & \textbf{Success} & \textbf{Failure} & \textbf{p50(ms)} & \textbf{p95(ms)} & \textbf{p99(ms)} \\
    \midrule
    L0 & warmup   & 48 & 100\% & 0\% & 2.96 & 3.1 & 7.6 \\
    L0 & attack   & 49 & 100\% & 0\% & 3.22 & 3.8 & 6 \\
    L0 & cooldown & 49 & 100\% & 0\% & 3.18 & 3.62 & 6.88 \\
    \midrule
    L3 & warmup   & 49 & 100\% & 0\% & 3 & 3.8 & 12.25 \\
    L3 & attack   & 49 & 55.6\% & 44.4\% & 1275.74 & 2002.2 & 2002.2 \\
    L3 & cooldown & 46 & 100\% & 0\% & 3.25 & 4.1 & 14.58 \\
    \bottomrule
    \end{tabular}
    }
    \caption{L0 vs.\ L3: HTTP success rate and censored latency.}

    \label{tab:ataques_t3_censored}
\end{table}

Table~\ref{tab:ataques_t3_censored} further details the aggregated behavior of the HTTP service under L3 across the \textit{warmup}, \textit{attack}, and \textit{cooldown} phases, considering success rate, failure rate, and censored latency. During \textit{warmup}, the service remains stable, sustaining 100\% success and low latency (p50 = 3\,ms and p95 = 3.8\,ms), which defines the nominal operating baseline. During attack phase, degradation becomes severe: the success rate drops to 56\%, while median latency rises to 1275.74\,ms and both p95 and p99 reach approximately 2002\,ms, consistent with the censorship threshold adopted in the experiment. This indicates that the attack compromises not only request availability but also the responsiveness of successfully completed requests, which operate close to the timeout limit.

\begin{figure}[!htb]
    \centering
    \resizebox{.95\columnwidth}{!}{
    \begin{minipage}{.5\textwidth}
        \centering
        \includegraphics[width=0.99\textwidth]{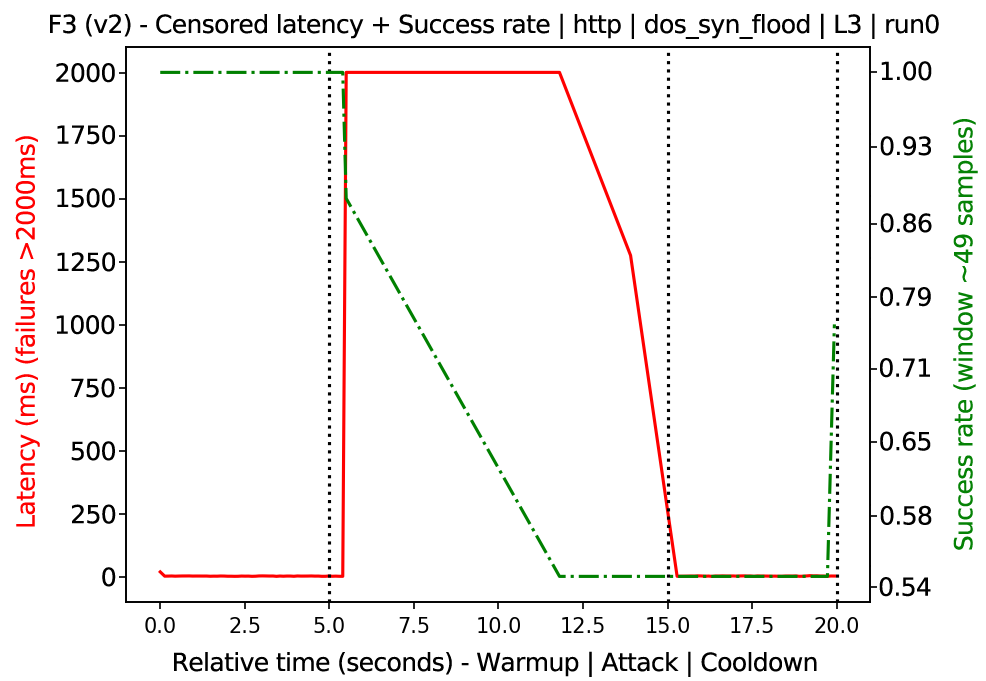}
        \caption{Censored latency versus success rate under HTTP SYN Flood DoS.}
        \label{fig:F3_censored_sr}
    \end{minipage}%
    \begin{minipage}{0.60\textwidth}
        \centering
        \includegraphics[width=0.99\textwidth]{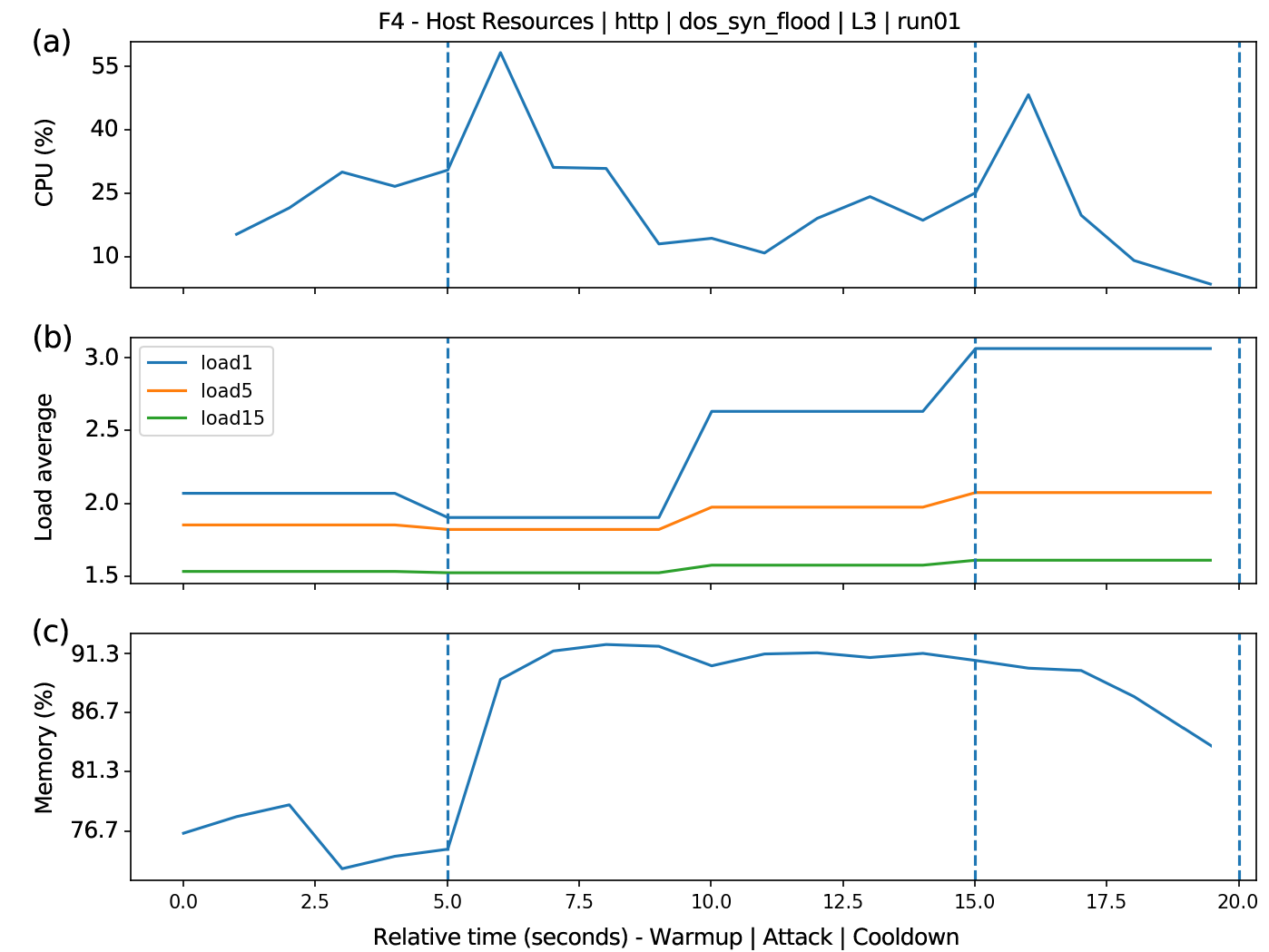}

        \caption{Target server resource usage.}

        \label{fig:F4_resources}
    \end{minipage}
    }
\end{figure}

After the attack interruption, the \textit{cooldown} phase shows full recovery, restoring 100\% success and near-baseline latency (p95 = 4.1\,ms). This behavior suggests that the observed degradation is strongly localized within the attack window, with no relevant persistence effects after interruption. Taken together, these results demonstrate that, at the most severe level, the \texttt{DoS SYN Flood} simultaneously compromises both service availability and responsiveness, whereas L0 preserves near-nominal behavior. The same trend is visually confirmed in Figure~\ref{fig:F3_censored_sr}.

In the illustrative results, we adopt a 5\,s \textit{warmup}, 10\,s continuous attack window, and 5\,s \textit{cooldown}. From the same experimental run, the framework also derives latency-versus-success metrics before, during, and after the attack, clearly capturing the degradation and temporary disruption of the HTTP service during the attack window. Requests with latency above 2000\,ms are treated as unsuccessful. Additional results for other attacks are available in the \textit{NetSecBed} repository\footnote{\url{https://github.com/ANONIMIZADO_PARA_REVISAO}}.

Figure~\ref{fig:F4_resources} further details the temporal evolution of three host-level resource metrics collected from the target server container during the 20\,s experiment. Figure~\ref{fig:F4_resources}(a) shows CPU utilization, which exhibits moderate oscillation during \textit{warmup} and a pronounced increase immediately after attack onset at 5\,s, reaching an early peak and a second elevation near the end of the attack window. Figure~\ref{fig:F4_resources}(b) presents system load averages (\textit{load1}, \textit{load5}, and \textit{load15}); as expected for a short experiment, \textit{load1} is the most sensitive and increases sharply during the attack, remaining elevated into the initial \textit{cooldown}, whereas \textit{load5} and \textit{load15} vary more smoothly due to their longer temporal windows. Figure~\ref{fig:F4_resources}(c) shows memory usage, which increases from 75--77\% before the attack to 89--91\% during the attack phase, with only partial recovery by the end of the observation window.

Taken together, these results show that the attack affects not only service-level availability and latency, but also imposes measurable internal pressure on the target server. CPU reacts immediately, system load remains elevated for longer, and memory usage does not instantly return to its pre-attack baseline. These host-level metrics therefore complement the evidence obtained from application probes and traffic captures, providing a clearer interpretation of both degradation mechanisms and recovery dynamics.

\section{Conclusion}
\label{sec:consideracoes_trabalhos_futuros}

This work presented \textit{NetSecBed}, a container-native testbed and methodological framework for reproducible cybersecurity experimentation that standardizes the full experimental lifecycle, enabling controlled scenario orchestration, traceable evidence generation (e.g., \textit{pcap}, logs, and extracted features), and systematic comparison across runs and scenarios, thereby advancing the state of the art in scalability, reproducibility, and continuous dataset generation for network security research. Experimental results further demonstrate its effectiveness in capturing and quantifying availability attacks on instrumented services: in the evaluated HTTP SYN Flood scenario, the service maintained 100\% success rate with p95 latency below 4\,ms during both \textit{warmup} and \textit{cooldown}, whereas the maximum intensity level (L3) reduced success to 55.6\% and increased p50 latency to over 1200\,ms during the attack window, followed by full recovery after interruption. These results confirm \textit{NetSecBed} as a robust, reproducible, and extensible evidence-generation framework, particularly well suited for controlled experimentation in IoT, IIoT, cyber-physical, and ubiquitous computing environments.

\bibliographystyle{plain}
\bibliography{sbc-template}

\end{document}